\documentclass[twocolumn,a4paper]{article}
\usepackage{bm,amsmath}
\usepackage{graphicx}
\usepackage{cite}
\usepackage{authblk}
\usepackage{hyperref}
\hypersetup{
	colorlinks=true,
	linkcolor=red,
	citecolor=blue
}
\usepackage[left=1.5cm,right=1.5cm,top=2cm,bottom=2cm]{geometry}

\begin{document}
\title{Steady photo-darkening of thulium alumino-silicate fibers pumped at $1.07~\mu m$: Quantitative effect of lanthanum, cerium and thulium. }

\author[1]{Jean-François Lupi}
\author[1]{ Manuel Vermillac}
\author[1,*]{Wilfried Blanc}
\author[1]{Franck Mady}
\author[1]{Mourad Benabdesselam}
\author[1]{Bernard Dussardier}
\author[2]{Daniel R. Neuville}

\affil[1]{Université Nice Sophia Antipolis, CNRS, Laboratoire de Physique de la Matière Condensée, UMR7336, Parc Valrose 06108 Nice, France}
\affil[2]{\textsc{\' e}quipe des géomatériaux , IPGP, COMUE Sorbonne Paris-Cité, CNRS UMR 7154, 1 rue Jussieu, 75005 Paris,France}

\twocolumn[
\begin{@twocolumnfalse}
\maketitle

\begin{abstract}
By pumping thulium-doped silica-based fibers at $\boldsymbol{1.07~\mu m}$, rapid generation of absorbing centers leads to photo-induced attenuation (PIA). This detrimental effect prevents exploiting laser emissions in the visible and near infrared. We report on the characterization of the PIA versus the fiber core composition, particularly the concentration of thulium ($\boldsymbol{Tm}$), lanthanum ($\boldsymbol{La}$) and cerium ($\boldsymbol{Ce}$) ions. We show that UV emision induced by $\boldsymbol{Tm}$-$\boldsymbol{Tm}$ energy transfers is the source of photo-darkening, and that lanthanum and cerium are efficient hardeners against PIA.
\end{abstract}
\bigskip
\end{@twocolumnfalse}]

Silica as a glass host for rare-earth (RE) doped lasers and amplifiers offers the best performances in terms of efficiency, power, reliability and cost effectiveness \cite{richardson2010high}. It is indeed chemically stable, mechanically robust and its optical transparency is high in the visible and the near-infrared (NIR) range up to $2~\mu m$. Some new applications of RE-doped silica materials require to develop optical transitions at wavelengths shorter than $0.85~\mu m$ \cite{leconte2015extended}. Applications of RE-transitions in the visible, or in uncovered spectral ranges in the NIR, are possible when using so called soft glasses, that are transparent in these ranges. In particular, glasses based on zirconium fluoride, like ZBLAN, have allowed many demonstrations of lasers and amplifiers from $0.45$ to $3.9~\mu m$ \cite{zhu2010high}. These glasses have a lower phonon-energy than silica: this allows high population inversion and efficient up-conversion pumping schemes. However, compared to silica, soft glasses are chemically and mechanically less stable, and are more expensive. Further, their threshold to optical damage is too low for the development of high power fiber applications.

\begin{figure}[htbp]
\centering
\includegraphics[width=\linewidth]{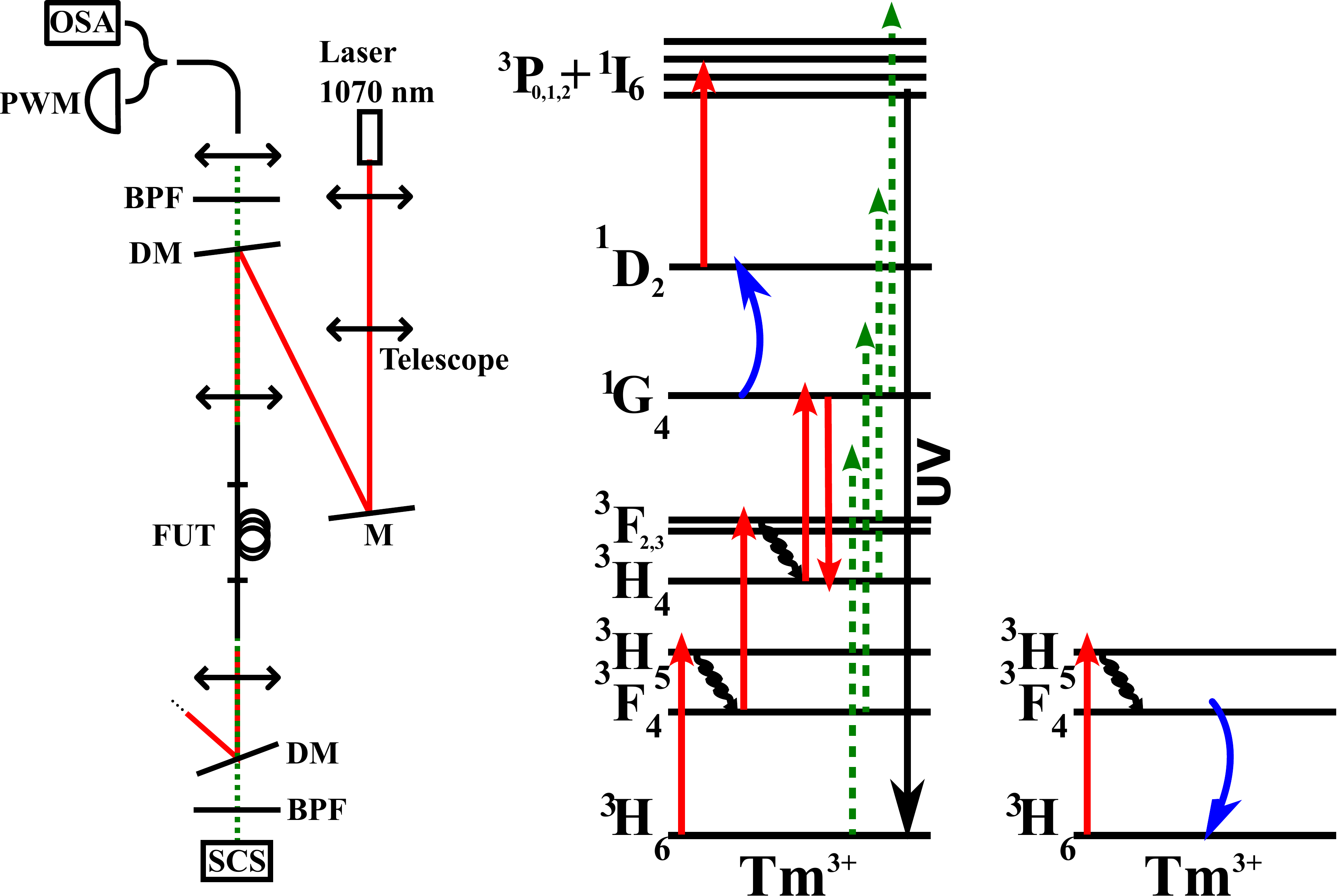}
\caption{(Left): Experimental setup for PIA measurement. M: dielectric mirror, FUT: fiber under test, BPF: band pass filter, DM: dichroic mirror, OSA: optical spectrum analyzer, SCS: supercontinuum source, PWM: power meter. (Right): Energy level diagram of $Tm^{3+}$: red: pump~induced transitions, blue: energy transfer, green: $550~nm$ probe.}
\label{manip}
\rule{\linewidth}{.5pt}
\end{figure}

Among the RE ions, thulium ($Tm^{3+}$) is particularly interesting because it offers many potential optical transitions spanning from $0.45$ to $1.9~\mu m$. Except for the current pumping scheme at $0.79~\mu m$ which provides efficient amplification around $1.9~\mu m$ through a down-conversion energy transfer process \cite{hanna1988continuous}, none of the other transitions are exploited in silica. An interesting up-conversion pumping scheme using a pump source at $1.07~\mu m$ provides up to seven laser emissions (spanning from $0.45$ to $1.9~\mu m$) that were implemented in ZBLAN fibers \cite{zhu2010high}. If one applies this scheme on a $Tm$-doped silica fiber, two phenomena will hamper amplification: fluorescence quenching by non-radiative decay and transparency degradation by photo-darkening \cite{broer1993highly}. The high phonon-energy of silica glass (as compared with fluorides) induces fast non-radiative decays from most energy levels of thulium, causing a strong reduction of their effective lifetime and hence the reachable population inversion, even under strong pumping.  In $Tm$-doped silica, the non-radiative decays from the $^3H_4$ excited level may be mitigated by highly co-doping with aluminum, promoting the amplification of the $0.8$ and $1.47~\mu m$ emission bands as well as upconversion to higher energy levels (see Figure \ref{manip})  \cite{blanc2008thulium,faure2007improvement,van1983nonradiative}.  Because the seeked applications necessitate high concentrations of aluminum and thulium, strong photodarkening is observed \cite{jetschke2013evidence,broer1993highly}. Besides, pumping of RE-doped optical fibers may cause the degradation of the transparency of both silica and fluoride glasses, especially when they are heavily doped \cite{engholm2009improved,broer1993highly,engholm2008preventing,litzkendorf2012study,jetschke2007equilibrium,jetschke2013evidence}. In the case of thulium-doped silica fiber pumped at $1.07~\mu m$, photo-darkening (or photo-induced attenuation, PIA) is particularly fast and intense, enough to prevent amplification in the extended-visible region ($0.45$-$0.9~\mu m$) \cite{broer1993highly}. The aim of this paper is to quantify the effects of the concentrations of $Tm$, $La$ and $Ce$ on the photo-darkening mechanism. The positive effects of reduction of photo-darkening have been already reported for ytterbium-doped fibers by co-doping with $Ce$ and explained by the ability of $Ce$ to exist both in 3+ and 4+ valence states \cite{engholm2009improved,unger2013optical}. $La$ is of interest in this study because it exists only in one valence state (3+) and is optically inactive (no absorption band). Here we show that the co-doping of $Tm$-doped silica fibers with either cerium or lanthanum causes a reduction of the PIA under $1.07~\mu m$ pumping. We propose a systematic study on the effect of $Tm$ and codopant concentrations on the PIA.

Fibers samples were prepared in our laboratory, using MCVD, solution doping and a drawing tower. The concentrations of $Al$, $Tm$, $La$ and $Ce$ were measured by electron probe micro analysis (EPMA) in the fibers. Three sets of samples were prepared. In each set, the concentration of only one ion varies (Table \ref{tab}).
 Attenuation spectra were measured using the cut back method on all samples. Taking the mode/dopant overlap factor into account for each fiber sample (typically 0.6 to 0.8), a linear correlation is obtained between attenuation values measured at $1190~nm$ (hump on the $^3H_5$ level ground state absorption band) and EPMA concentration, leading to a correspondence of $1~dB/m$ at $1190~nm$ for $\sim$ $18~ppm.at$ of $Tm^{3+}$.
 
\begin{table}[htbp]
\centering
\caption{\bf Concentrations (ppm.at of all elements incl. oxygen)}
\begin{tabular}{c||cccc}
\hline
Series & $Tm$ & $La$ & $Ce$ & $Al$\\
\hline
‘$Tm$’ & $0$-$600$ & - & - & $\sim 8000$ \\
‘$La$’ & $190\pm 30$ & $0$-$7000$ & - & $\sim 8000$ \\
‘$Ce$’ & $260\pm 40$ & - & $0$-$1300$ & $\sim 8000$ \\
\hline
\end{tabular}
  \label{tab}
\end{table}

Figure \ref{manip} describes the experimental setup used to measure the PIA. The pump laser at $1.07~\mu m$ is a continuous wave ytterbium-doped fiber laser. It is coupled into a commercial passive fiber (Corning, HI-1060) through a telescope, a high reflection dielectric mirror, a dichroic mirror and an aspherical lens. The dichroic mirror is highly reflective at $1.07~\mu m$ and transparent from $0.55$ to $1.0~\mu m$. The input fiber end is cleaved at right angle. The fiber under test (FUT) samples are spliced to the input fiber. The typical length of samples is about 2 cm, short enough to neglect the pump depletion. Another passive HI-1060 fiber is spliced at the other end of FUT. A super continuum source (SCS) is coupled into the passive fiber in the counter-propagative direction relative to the pump beam, using two metallic mirrors (not represented) and an aspherical lens. The remaining pump is rejected by a second dichroic mirror, so that the SCS is protected against laser damage. The infra-red part of the SCS (> 1 µm) is filtered out by a short-pass filter before the coupling into the fiber. When necessary, two band pass filters are placed before and after the FUT. The bandwidth of band pass filters is $2~nm$ at the central wavelength $550~nm$. Their extinction ratio is higher than $40~dB$. Next, the green $2~nm$ broad beam is called the probe. 

The injected pump is measured by a power meter before the splicing of the FUT. Spectrally and time resolved measurements of the PIA were performed. The PIA spectra were measured from $550$ to $1000~nm$, at room temperature. To this aim, the band pass filters were removed and the OSA (Anritsu, MS9030A, 350-1750 nm) was used. A reference spectrum is recorded before the pump is turned on, and a second one is recorded just after the pump is turned off. In this configuration, the power of the SCS beam coupled into the FUT is typically less than $1~mW$. To minimize the photo-bleaching which may be induced by the probe, the SCS is turned off during pump exposure. Just after the pump laser is turned off, the OSA is set to record the spectrum within 10 seconds only. The PIA spectrum (in $dB/m$) is computed from transmission spectra and normalized by the FUT length ($\sim 2~cm$). The  time-resolved PIA measurement was recorded at $550~nm$ while pumping at $1070~nm$. This probe wavelength was selected because the PIA is strong whereas there is no absorption band of $Tm$ ions at $550~nm$. The band pass filters are placed, the SCS is turned on, and the PWM (Si photo-diode) is used to continuously measure the transmitted probe while pumping. The recording is continued after the laser is turned off, in order to check that almost no bleaching is caused by the probe. 

\begin{figure}[htbp]
\centering
\includegraphics[width=\linewidth]{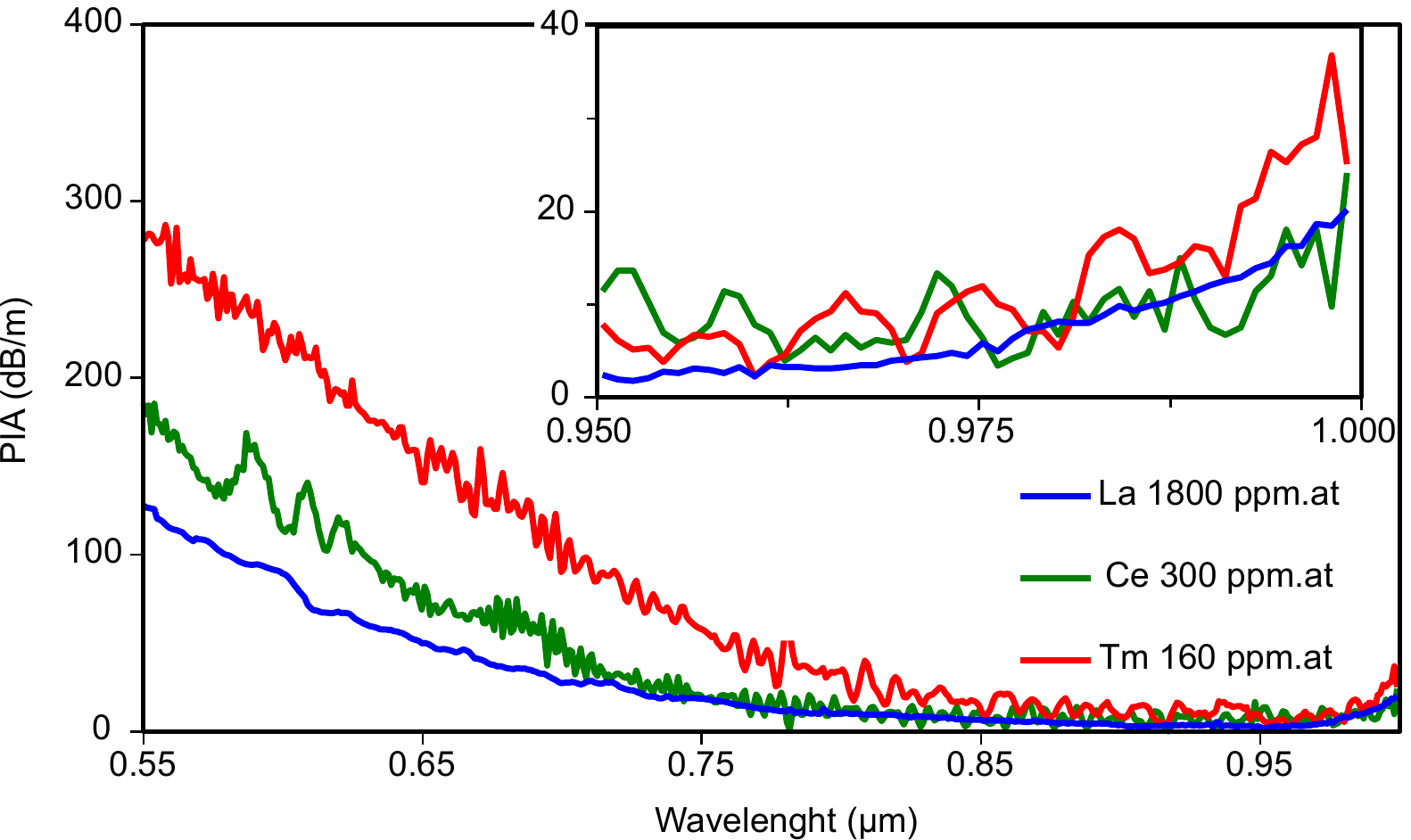}
\caption{PIA spectra from selected samples after $1.07~\mu m$ laser exposure for $30~min$ (pump power $750~mW$). The resolution is $10~nm$ for the ‘$La$’ samples and $1~nm$ for the ‘$Ce$’ and ‘$Tm$’ samples. Acquisition time is 10 seconds. Inset: Zoom on the near infra-red region.}
\label{spectrum}
\rule{\linewidth}{.5pt}
\end{figure}

Figure \ref{spectrum} shows the PIA spectra obtained by measuring of the power of the transmitted SCS beam before and after laser exposure at $750~mW$ during 30 minutes, for representative samples studied in this work: ‘$La~1800~ppm.at$’, ‘$Ce~300~ppm.at$’ and ‘$Tm~160~ppm.at$’.

By pumping the fiber at $1.07~\mu m$, two absorption (PIA) appear, a broad and strong absorption in the visible ($\lambda$ < $900~nm$) and a weak one in the near infra-red ($\lambda$ > $950~nm$) (Figure \ref{spectrum}). The latter band is not fully scanned because of the dichroic mirrors cut-off above $1000~nm$. Cerium ions (both 3+ and 4+) absorb at wavelengths shorter than $400~nm$ \cite{hamilton1989optical,dorenbos2003systematic,paul1976ultraviolet,raukas1996luminescence,johnston1965oxidation} and $La$ is optically inactive. Hence, these two absorptions cannot be attributed to direct absorptions by $La$ or $Ce$. Therefore, PIA absorption are mainly related to the presence of both $Tm$ and $Al$ ions. The visible band has been already reported in rare-earth doped alumina-silica fiber and it is related to the presence of $Al$ induced absorbing centers \cite{engholm2008preventing}. On the contrary, the infra-red absorption band has not been reported in non thulium-doped alumina silicates \cite{engholm2008preventing,mebrouk2014experimental}, indicating that it could be related to $Tm$ ions. An absorption band in near infra-red has been reported in $CaF_2$ and was attributed to 4f-4f transitions of $Tm^{2+}$ \cite{kiss1962energy}. This suggests that during the photo-darkening process, charge transfers would generate color centers, as in $Yb$-doped fibers, and in addition would reduce $Tm^{3+}$ into $Tm^{2+}$, as already observed with $Er$ ions \cite{mebrouk2014experimental}. 4f-5d absorption bands of $Tm^{2+}$ are also expected and may also contribute to the visible band \cite{kiss1962energy}.

\begin{figure}[htbp]
\centering
\includegraphics[width=\linewidth]{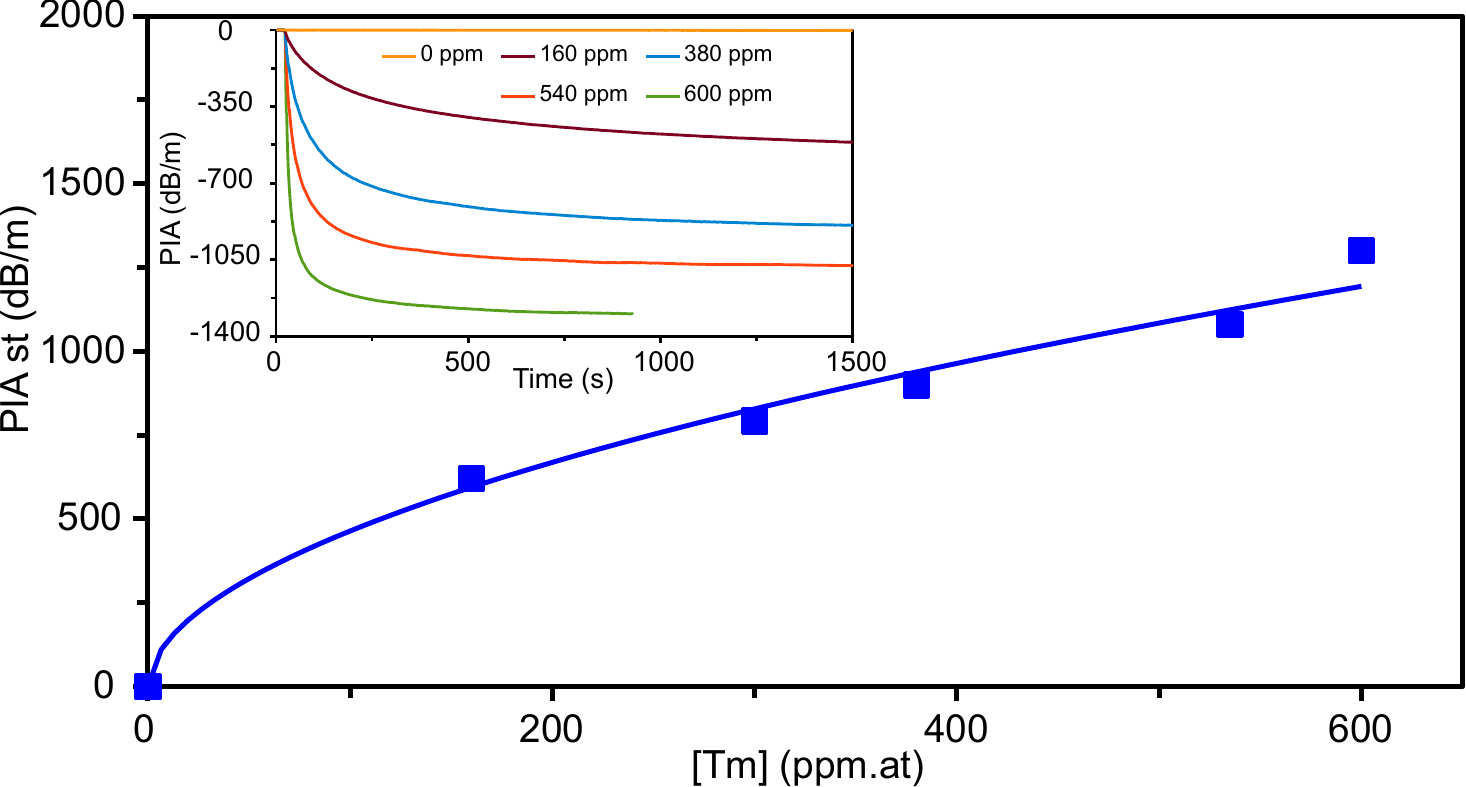}
\caption{Steady state PIA (PIAst) versus thulium concentration in the Al-Tm-codoped ‘$Tm$’ series, recorded at $550~nm$. Pump power: $1~W$. Squares: experimental data, solid line: power-law fit with exponent = $\tfrac{1}{2}$. Inset: corresponding temporal curves.}
\label{fig_tm}
\rule{\linewidth}{.5pt}
\end{figure}

Figure \ref{fig_tm} shows the steady PIA (PIAst) versus the thulium concentration for the ‘$Tm$’ series. The inset shows the temporal curves. The PIA converges to a steady state (PIAst) as described in \cite{jetschke2007equilibrium}. The PIAst value for the $Tm$ series increases from $0$ to $1300~dB/m$ when the thulium concentration increases from $0$ to $600~ppm.at$. The best candidate as a function to fit the results is a power-law, with an exponent equals to $\tfrac{1}{2}$ for the dependence on $Tm$ concentration (solid line in Figure \ref{fig_tm}). The same trend was observed in Yb-doped silica with varying $Tm$ concentration and probe light at $633~nm$ \cite{jetschke2013evidence}.

\begin{figure}[htbp]
\centering
\includegraphics[width=\linewidth]{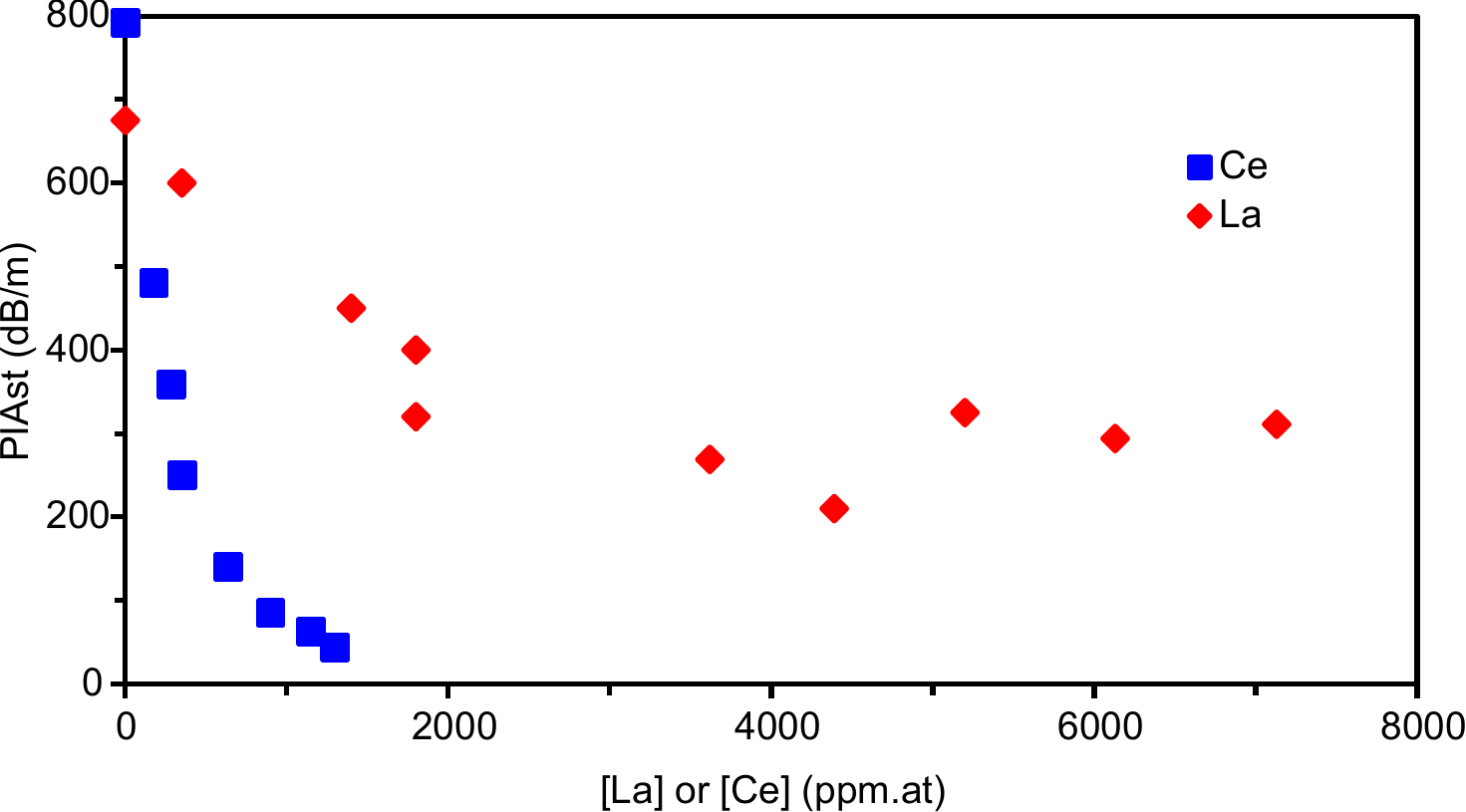}
\caption{Steady state PIA (PIAst) versus cerium concentration in the ‘$Ce$’ (Ce-Al-Tm) series (Blue) and versus lanthanum concentration in the ‘$La$’ (La-Al-Tm) series (Red), recorded at $550~nm$. Pump power: $1~W$. Squares: experimental data.}
\label{fig}
\rule{\linewidth}{.5pt}
\end{figure}

PIAst was measured in both ‘$La$’ and ‘$Ce$’ series. Its behaviour versus the concentration of lanthanum in the ‘$La$’ series is shown in Figure \ref{fig}. The point at $0~ppm.at~La$ is extracted from Figure \ref{fig_tm}. When $La$ concentration increases up to $4000~ppm.at$, the PIAst reduces from $675$ to les than $300~dB/m$. Above $4000~ppm.at$ PIAst is almost constant.

The PIAst versus the concentration of cerium in the ‘$Ce$’ series is shown in Figure \ref{fig}. Cerium reduces the PIAst from $795$ to $44~dB/m$ when the $Ce$ concentration varies from $0$ to $1300~ppm.at$.

\begin{figure}[htbp]
\centering
\includegraphics[width=\linewidth]{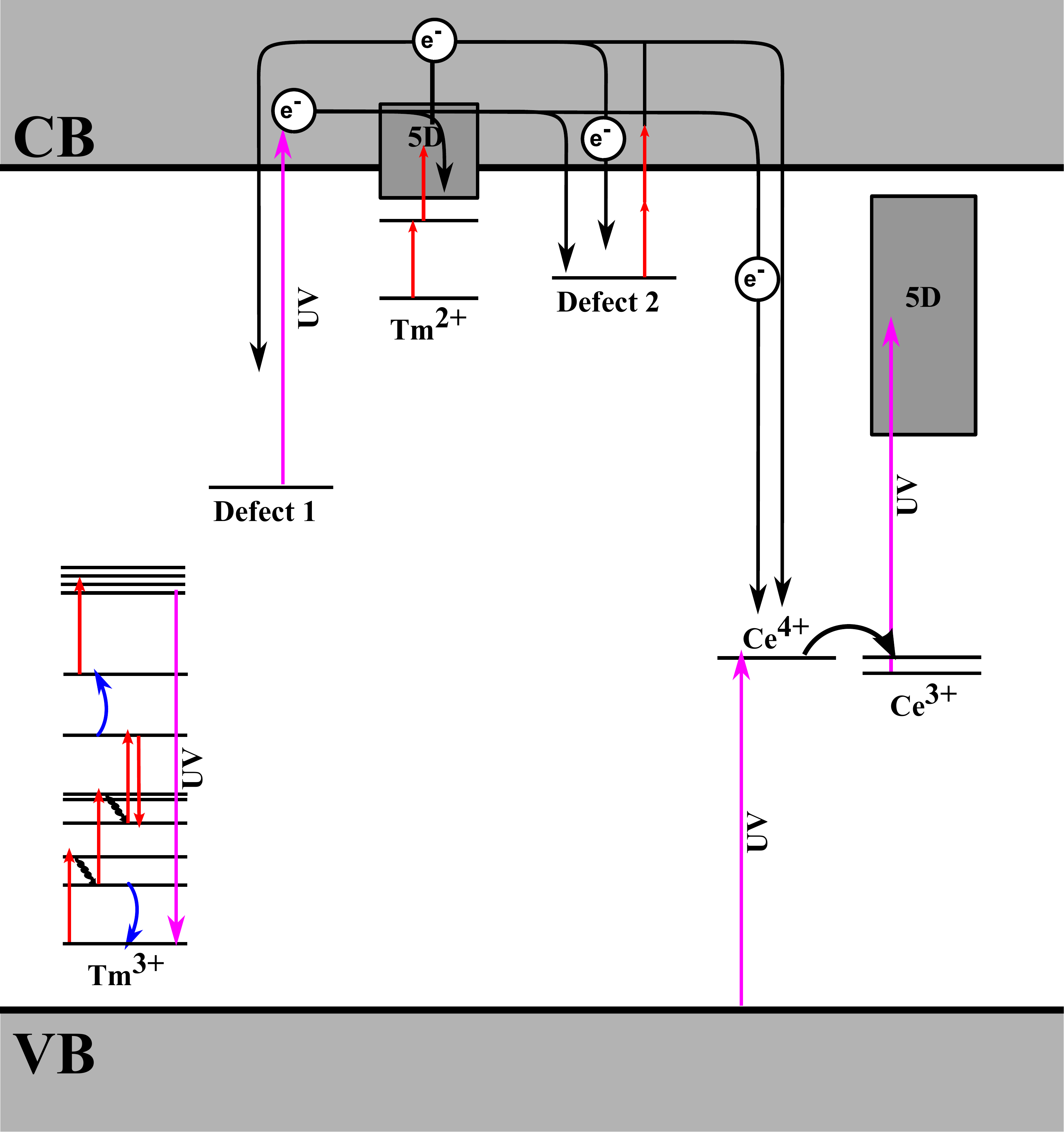}
\caption{Schematic representation of photo-darkening process, for $Tm$ pumped at $1.07~\mu m$. An animated sequence of these mechanisms is available as supplementary material (see Visualization 1).}
\label{fig_PD}
\rule{\linewidth}{.5pt}
\end{figure}

Based on these results, we propose the following interpretation. Clusters of rare-earth, defects in glass matrix and generation of UV photons by energy transfers are usually invoked as the cause of photo-darkening \cite{engholm2008preventing}. Such mechanisms may be identified to explain the behaviour of PIAst when the $Tm$ content increases. It is well known that when the $Tm$ concentration increases, clusters tend to form, even with aluminum co-doping \cite{monteil2004clustering,simpson2006energy}. By pumping the fiber at $1.07~\mu m$, $Tm^{3+}$ is excited up to the $^1G_4$ level (Figure \ref{fig_PD}). Then, the  $Tm$-clusters promotes the energy transfer between $Tm$ ions to populate the $^1D_2$ level. The population of this level is charaterized by emissions bands at 520, 540, 660 and 740 nm (not shown). Thanks to the absorption of a fifth $1.07~\mu m$ photon, higher levels are reached ($^1I_6, ^3P_{0,1,2}$) (Figures \ref{manip} and \ref{fig_PD}), leading to emssion of UV photon \cite{paschotta1997characterization}. This UV photon can ionize defects in alumino-silicates (e.g. AlOHC, Al-E$\prime$, NBOHC, ODC \cite{skuja1998optically}), leading to the promotion of an electron in the conduction band.  This electron may recombine on a $Tm^{3+}$ ion to form a $Tm^{2+}$ ion or be trapped by a precursor center and form a color center. These steps are related to the formation of PIA. Steady state is reached because $Tm^{2+}$ ions and ``\textit{Defect 2}'' can be ionized by absorbing two pump photons (Figure \ref{fig_PD}). All these steps are depicted in an animated sequence available as supplementary material (see Visualization 1).

As $La^{3+}$ ion is optically inactive, the main influence of this ion should be related to structural effects. When the RE ion is added into alumino-silicate glasses, it acts as a charge compensator if $\tfrac{[Al_2O_3]}{[RE_2O_3]}~\leq~3$ \cite{florian200727al}. As a consequence, RE is located preferentially close to the tetra-coordinated $Al^{IV}$ species \cite{florian200727al}. $La$ and $Tm$ compete to be located close to $Al^{IV}$ species. Therefore, $La$ tends to reduce the probability of $Tm$-cluster formation and hence the emission of UV photons. The same role for $La$ has been proposed in erbium-lanthanum-doped silica fibers \cite{philipsen1999observation}. For $La$ concentration above $4000~ppm.at$, the influence of $La$ changes. As $Al$ content is about $8000~ppm.at$, these concentrations of $La$ are above the threshold requested for the charge compensation (typically $2500~ppm.at~La$ in our case). Then, $La$ acts as network modifier and depolymerize the silica network. This depolymerization may lead to the formation of new color-centers which may absorb in the visible as already observed with $Ce^{3+}$ ions \cite{demos2014dynamics}. A change of $Tm^{3+}$ ions environment is also possible. For example, a higher content of $La$ in the $Tm$ environment may decrease the local phonon energy ($ E_{p}(La_2O_3) = 400cm^{-1}$, $ E_{p}(Al_2O_3) = 870cm^{-1}$,$ E_{p}(SiO_2) = 1100cm^{-1}$), then enhance the population in $^1G_4$ level, and therefore the energy transfer mechanism \cite{blanc2008thulium,faure2007improvement,van1983nonradiative}.

Cerium is more efficient than lanthanum to mitigate photo-darkening. Indeed, at low concentration, $Ce$ and $La$ have a reducing rate of $-1.6~dB/m/ppm.at$ and $-0.16~dB/m/ppm.at$, respectively. $Ce$ is ten times more efficient than $La$ at low concentration. However, the role of $Ce$ is more complex to discuss than $La$ because it has two valence states (4+ and 3+) and both of them could combine structural and optical effects. The characterization and the control of the redox is a major issue, particularly in glasses. It depends on many factors such as the ion concentration, the overall composition, the temperature at which the glass was melted, the furnace atmosphere (oxidizing or reducing) \cite{densem1938equilibrium}. Even in oxidizing conditions, the redox of cerium is not clearly established. Previous reports support the predominance of either $Ce^{4+}$ or $Ce^{3+}$, but their conclusions are highly dependant on fabrication conditions \cite{fasoli2009effect,johnston1965oxidation,unger2013optical}. When cerium is present as $Ce^{4+}$, it would be efficient to compensate the charge of $Al^{IV}$ species ($\tfrac{[Al_2O_3]}{[CeO_2]}~\leq~4$  and $\tfrac{[Al_2O_3]}{[Ce_2O_3]}~\leq~3$). This could be a first explanation of the highest rate of photodarkening mitigation by $Ce$ as the threshold for charge compensation should be around $2000~ppm.at~Ce$. $Ce^{4+}$ has a charge transfer absorption band peaking at $260~nm$ and $Ce^{3+}$ has a $4f$-$5d$ transition at $330~nm$ \cite{fasoli2009effect}. therefore both valence states of $Ce$ can absorb UV emitted by $Tm$ (mitigating the ionization of defects and releasing electrons in the conduction band (Figure \ref{fig_PD} and Visualization 1). $Ce^{4+}$ can also trap the electron released in the conduction band, thus reducing the formation of $Tm^{2+}$ and ``\textit{Defect 2}''.

In summary, series of MCVD-made fibers were fabricated containing various concentrations of thulium, lanthanum and cerium. Spectral and time-resolved measurements of photo-induced absorption were performed under high power pumping at $1.07~\mu m$ wavelength. PIA spectra are all similar, with a strong visible and a weak near-infra-red broad absorption bands. They suggest a charge transfer during photo-darkening, transforming $Tm^{3+}$ into $Tm^{2+}$. PIA increases with $Tm$ concentration. The lanthanum and the cerium co-doping are found to behave as hardeners against photo-darkening. Those effects are tentatively assigned to their structural and optical effects. It is found that cerium is a very good additive for reducing photo-darkening because 95$\%$ of this detrimental effect is bleached by adding $1300~ppm.at$ of $Ce$. It is also found that $La$ can reduce photo-darkening by 70$\%$. We believe that this quantitative study contributes to the understanding of photo-degradation of Tm-doped fibers under $1.07~\mu m$ pumping with the aim of developing of powerful fiber amplifiers at wavelengths shorter than $0.85~\mu m$. First, the 1300-ppm $Ce$ sample PIAst is 10 times lower than Tm absorption in the NIR, so amplification would be achievable along the  $^1G_4 \rightarrow ^3F_4$ transition at $0.785~\mu m$. Further, increasing of the cerium concentration is possible without loss of glass quality up to $0.55\%$ of non-oxygen elements at least (>~10 times more than in our samples) \cite{engholm2009improved}. This would further lower PIAst levels, and hence allow for amplification at shorter wavelengths. 

Funding. Université Nice Sophia Antipolis, Centre National de la Recherche Scientifique, Agence Nationale de la Recherche (ANR-14-CE07-0016-01, Nice-DREAM).

Acknowledgements. We thank S. Trzesien and M. Ude (LPMC, Nice) for the fabrication of the samples, M. Fialin (IPGP Camparis) for EPMA measurements and B. Gay-Para (LPMC, Nice) for the preparation of the animated sequence. 
\bigskip

\bibliographystyle{unsrt}
{\scriptsize  \bibliography{sample}}

\end{document}